
\input harvmac
\def\figflag{I}
\def\psheader#1{}


\font\blackboard=msbm10 \font\blackboards=msbm7
\font\blackboardss=msbm5
\newfam\black
\textfont\black=\blackboard
\scriptfont\black=\blackboards
\scriptscriptfont\black=\blackboardss
\def\blackb#1{{\fam\black\relax#1}}



%
\def\BC{{\blackb C}} 
 
\def\BZ{{\blackb Z}} 
\def\BP{{\blackb P}}

\def\Bid{{\mathchoice {\rm {1\mskip-4.5mu l}} {\rm
{1\mskip-4.5mu l}} {\rm {1\mskip-3.8mu l}} {\rm {1\mskip-4.3mu l}}}}

%
\font\mathbold=cmmib10 \font\mathbolds=cmmib7
\font\mathboldss=cmmib5
\newfam\mbold
\textfont\mbold=\mathbold
\scriptfont\mbold=\mathbolds
\scriptscriptfont\mbold=\mathboldss
\def\bi{\fam\mbold\relax}


%
\font\gothic=eufm10 \font\gothics=eufm7
\font\gothicss=eufm5
\newfam\gothi
\textfont\gothi=\gothic
\scriptfont\gothi=\gothics
\scriptscriptfont\gothi=\gothicss

%

\def\tfig#1{Fig.~\the\figno\xdef#1{Fig.~\the\figno}\global\advance\figno by1}
\def\figI{I}
%
\newdimen\tempszb \newdimen\tempszc \newdimen\tempszd \newdimen\tempsze
\ifx\figflag\figI
\input epsf
%
\def\epsfsize#1#2{\expandafter\epsfxsize{
 \tempszb=#1 \tempszd=#2 \tempsze=\epsfxsize
     \tempszc=\tempszb \divide\tempszc\tempszd
     \tempsze=\epsfysize \multiply\tempsze\tempszc
     \multiply\tempszc\tempszd \advance\tempszb-\tempszc
     \tempszc=\epsfysize
     \loop \advance\tempszb\tempszb \divide\tempszc 2
     \ifnum\tempszc>0
        \ifnum\tempszb<\tempszd\else
           \advance\tempszb-\tempszd \advance\tempsze\tempszc \fi
     \repeat
\ifnum\tempsze>\hsize\global\epsfxsize=\hsize\global\epsfysize=0pt\else\fi}}
\epsfverbosetrue
\psheader{fig3.pro}       
\fi
%

%
%
%
%

\def\ifigure#1#2#3#4{
\midinsert
\vbox to #4truein{\ifx\figflag\figI
\vfil\centerline{\epsfysize=#4truein\epsfbox{#3}}\fi}
\baselineskip=12pt
\narrower\narrower\noindent{\bf #1:} #2
\endinsert
}
%
%
\def\ifigures#1#2#3#4#5#6#7#8{
\midinsert
\centerline{
\hbox{\vbox{
\divide\hsize by 2
\vbox to #4truein{\ifx\figflag\figI
\vfil\centerline{\epsfysize=#4truein\epsfbox{#3}}\fi}
\baselineskip=12pt
\narrower\narrower\noindent{\bf #1:} #2
}}\qquad
\hbox{\vbox{
\divide\hsize by 2
\vbox to #8truein{\ifx\figflag\figI
\vfil\centerline{\epsfysize=#8truein\epsfbox{#7}}\fi}
\baselineskip=12pt
\narrower\narrower\noindent{\bf #5:} #6
}}}
\endinsert
}


\def\appendix#1#2{\global\meqno=1\global\subsecno=0\xdef\secsym{\hbox{#1:}}
\bigbreak\bigskip\noindent{\bf Appendix #1: #2}\message{(#1: #2)}
\writetoca{Appendix {#1:} {#2}}\par\nobreak\medskip\nobreak}

\def\ket#1{| #1 \rangle}         
\def\melt#1#2#3{\langle #1 \mid #2 \mid #3\rangle} 
%

\def\fourpt{\hbox{{$\rangle \kern-.25em \langle$}}} 
\def\tree{\hbox{{$\rangle \kern-.5em - \kern-.5em \langle$}}}
\def\ib{{\bar \imath}} 

\def\H#1#2{{\rm H}^{#1}(#2)} 
\def\CM{{\cal M}} \def\CW{{\cal W}}

\def\CS{{\cal S}}

\def\ET{{\rm End({\it T})}}
\def\ex#1{{\rm e}^{#1}}                 

\def\cp#1{{\BC{\rm P}^{#1}}}

\def\sdp{{\blackb n}}

\def\Ka{K\"ahler}

\def\LG{Landau-Ginzburg}
\def\sm{$\sigma$-model}

\def\ql{{\bi q}}
\def\qr{\overline{\bi q}}

\def\mm#1#2#3#4#5{\left({\textstyle #1 {#2\atop #4}{#3\atop #5}}\right)}
\long\def\optional#1{}

\noblackbox

\lref\DK{J. Distler and S. Kachru, ``(0,2) Landau-Ginzburg Theory,''
{\it Nucl. Phys.} {\bf B413} (1994) 213, {\tt hep-th/9309110}.}

\Title{\vbox{\hbox{PUPT--1465}\hbox{\tt hep-th@xxx/9406090}}}
{\vbox{\centerline{Singlet Couplings and (0,2) Models$^\star$}
}}

\centerline{Jacques Distler and Shamit Kachru$^\dagger$}\smallskip
\centerline{Joseph Henry Laboratories}
\centerline{Princeton University}
\centerline{Princeton, NJ \ 08544 \ USA}
\bigskip

\footnote{}{{\parindent=-5pt\par $\dagger$Address
after June 30: Department of Physics, Harvard University, Cambridge, MA
02138.}}
\footnote{}{{\parindent=-5pt\par $\star$
\vtop{
\hbox{Email: {\tt distler@puhep1.princeton.edu}, {\tt
kachru@puhep1.princeton.edu} .}
\hbox{Research supported by NSF grant PHY90-21984, and the
A.~P.~Sloan Foundation.}
     }     }}

We use the quantum symmetries present in string compactification on
Landau-Ginzburg orbifolds to
prove the existence of a large class of
exactly marginal (0,2) deformations
of (2,2) superconformal theories.
Analogous methods apply to the more general (0,2) models introduced in
\DK, lending further credence to the fact that the corresponding \LG\
models represent bona-fide (0,2) SCFTs.
We also use the large symmetry groups which arise
when the worldsheet superpotential is turned off to
constrain the dependence of
certain correlation functions on the
untwisted moduli. This allows us to approach the problem of what happens when
one tries to deform away from the \LG\ point.
In particular, we find that the masses and three-point
couplings of the massless $E_{6}$ singlets
related to ${\rm H^{1}}(\ET)$ vanish at all points in the
quintic \Ka\ moduli space.
Putting these results together,
and invoking some plausible dynamical assumptions
about the corresponding linear \sm s, we show that one can deform
these \LG\ theories to arbitrary values of the \Ka\ moduli.

\Date{June 1994}                 

\lref\Sei{N. Seiberg, ``Exact Results on the Space of Vacua of
Four-Dimensional SUSY Gauge Theories,'' Rutgers preprint, {\tt
hep-th/9402044}.}
\lref\Gepner{D. Gepner, ``String Theory on Calabi-Yau Manifolds: The
Three Generation Case,'' Princeton preprint, December 1987.}
\lref\Schimmrigk{R. Schimmrigk, ``A New Construction of a
Three-Generation Calabi-Yau Manifold,'' {\it Phys. Lett.} {\bf 193B}
(1987) 175.}
\lref\Leigh{M. Dine, R.G. Leigh, and D.A. MacIntire, ``Discrete
Gauge Anomalies in String Theory,'' Santa Cruz preprint, {\tt
hep-th/9307152}. }
\lref\WitMin{E. Witten, ``On the Landau-Ginzburg Description of N=2 Minimal
Models,'' IAS preprint, {\tt hep-th/9304026}.}
\lref\Zam{A.B. Zamolodchikov, ``Irreversibility of the Flux of the
Renormalization Group in a 2d Field Theory,'' {\it JETP Lett.} {\bf 43}
(1986) 730.}

\lref\AspSmall{P. Aspinwall, B. Greene, and D. Morrison, ``Measuring
Small Distances in N=2 Sigma Models,'' IAS preprint,
{\tt hep-th/9311042}.}

\lref\SpecialGeo{B. de Wit, P. Lauwers and A. van Pr\oe yen, Nucl. Phys. {\bf
B255} (1985) 569.}
\lref\ranspec{S. Cecotti, S. Ferrara and L. Girardello, Int. J. Mod. Phys.
{\bf A4} (1989) 2475\semi
L. Castellani, R. D'Auria and S. Ferrara, Class. Quantum Grav. {\bf 7} (1990)
1767.}
\lref\CandMod{P. Candelas and X. De la Ossa, ``Moduli
Space of Calabi-Yau Manifolds,'' {\it Nucl. Phys.} {\bf B355} (1991)
455.}
\lref\OldEd{E. Witten, ``Symmetry Breaking Patterns in Superstring
Models,'' {\it Nucl. Phys.} {\bf B258} (1985) 75.}
\lref\DKL{L. Dixon, V. Kaplunovsky and J. Louis, Nucl. Phys. {\bf B329}
(1990) 27.}
 \lref\StromSpec{A. Strominger, ``Special Geometry,''
 {\it Comm. Math. Phys.} {\bf 133}
(1990) 163.}
\lref\PerStrom{V. Periwal and A. Strominger, Phys. Lett. {\bf B335} (1990)
261.}
\lref\TopAntiTop{S. Cecotti and C. Vafa, Nucl. Phys. {\bf B367} (1991) 359.}
\lref\AandB{E. Witten, in ``Proceedings of the Conference on Mirror Symmetry",
MSRI (1991).}
\lref\LVW{W. Lerche, C. Vafa and N. Warner, Nucl. Phys. {\bf B324} (1989)
427.} 
\lref\twisted{E. Witten. Comm. Math. Phys. {\bf 118} (1988) 411\semi
E. Witten, Nucl. Phys. {\bf B340} (1990) 281\semi
T. Eguchi and S.-K. Yang , Mod. Phys. Lett. {\bf A5} (1990) 1693.}
 \lref\GVW{B. Greene, C. Vafa and N. Warner,
``Calabi-Yau Manifolds and Renormalization Group Flows,''
 {\it Nucl. Phys.} {\bf B324} (1989)
371.}
\lref\Grisaru{M. Grisaru, A. Van de Ven and D. Zanon, Phys. Lett. {\bf 173B}
(1986) 423.}
\lref\SeiNat{N. Seiberg, ``Naturalness Versus Supersymmetric
Non-renormalization Theorems,'' {\it Phys. Lett.} {\bf B318} (1993)
318, {\tt hep-ph/9309335}.}

\lref\DSWW {M. Dine, N. Seiberg, X.G. Wen and E. Witten,
``Non-Perturbative Effects on the String World Sheet I,'' {\it Nucl.
Phys.}~{\bf B278} (1986) 769, ``Non-Perturbative Effects on the String
World Sheet II,'' {\it Nucl. Phys.}~{\bf B289} (1987) 319. }
\lref\DixonRev{L. Dixon, in ``Proceedings of the 1987 ICTP Summer Workshop in
High Energy Physics and Cosmology", ed G. Furlan, {\it et. al.}}
\lref\Exact{J. Distler and B. Greene,
``Some Exact Results on the Superpotential from Calabi-Yau
Compactifications,''
{\it Nucl. Phys.}
 {\bf B309} (1988) 295.}
\lref\twozero{J. Distler and B. Greene,
``Aspects of (2,0) String Compactifications,''
{\it Nucl. Phys.} {\bf B304} (1988) 1.}
\lref\AspMor{P. Aspinwall and D. Morrison, ``Topological Field Theory and
Rational Curves," Comm. Math. Phys. {\bf 151} (1993) 245.}
\lref\CandMir{P. Candelas, X. de la Ossa, P. Green and L. Parkes,
``A Pair of Calabi-Yau Manifolds as an Exactly Soluble Superconformal
Theory,''
{\it Nucl.
Phys.} {\bf B359} (1991) 21.}
\lref\Kutconf{D. Kutasov, ``Geometry on the space of conformal field
theories and contact terms,'' Phys. Lett. {\bf B220} (1989) 153.}%
\lref\GrSei{M. Green and N. Seiberg, ``Contact interactions in
superstring theory," Nucl. Phys. {\bf B299} (1988) 559.}%
\lref\WilZee{F. Wilczek and A. Zee, Phys. Rev. Lett. {\bf 52} (1984) 2111.}
\lref\Banks{T. Banks, L. Dixon,
D.Friedan and E. Martinec, ``Phenomenology and Conformal Field Theory or
Can String Theory Predict the Weak Mixing Angle?,'' {\it Nucl. Phys.}
{\bf B299} (1988) 613.}
\lref\Hodge{P. Griffiths, Periods of Integrals on Algebraic manifolds I,II,
Am. J. Math. {\bf 90} (1970) 568,805\semi
R. Bryant and P. Griffiths, in ``Progress in Mathematics {\bf 36}"
(Birkh\"auser, 1983) 77.}
\lref\Odake{S. Odake, Mod. Phys. Lett. {\bf A4} (1989) 557\semi
S. Odake, Int. Jour. Mod. Phys. {\bf A5} (1990) 897\semi
T. Eguchi, H. Ooguri, A. Taormina and S-K. Yang, Nucl. Phys. {\bf B315}
(1989) 193.}
\lref\GSW{M. Green, J. Schwarz and E. Witten, ``Superstring theory, vol. II"
(Cambridge University Press, 1987).}
\lref\CHSW{P. Candelas, G. Horowitz, A. Strominger and E. Witten,
``Vacuum Configurations for Superstrings,'' {\it Nucl. Phys.} {\bf
B258} (1985) 46.}
\lref\NR{L. Alvarez-Gaum\'e, S. Coleman and P. Ginsparg, Comm. Math. Phys.
{\bf 103} (1986) 423.}
\lref\manin{D. Leites, ``Introduction to the theory of supermanifolds", Russ.
Math. Surveys {\bf 35} (1980) 3\semi
Yu. Manin, ``Gauge field theory and complex geometry", (Springer, 1988).}
\lref\Reviews{
J. Schwarz, ``Superconformal symmetry in string theory", lectures
at the 1988 Banff Summer Institute on Particles and Fields (1988)\semi
D. Gepner, ``Lectures on N=2 string theory",
lectures at the 1989 Trieste Spring School (1989)\semi
N. Warner, ``Lectures on N=2 superconformal theories and singularity theory",
lectures at the 1989 Trieste Spring School (1989)\semi
B. Greene, ``Lectures on string theory in four dimensions",
lectures at the 1990 Trieste Spring School (1990)\semi
S. Yau (editor), ``Essays in Mirror Manifolds",  Proceedings of the Conference
on Mirror Symmetry, MSRI (International Press, 1992).
}
\lref\Trieste{J. Distler, ``Notes on N=2 $\sigma$-models," lectures at the
1992 Trieste Spring School (1992), {\tt hep-th/9212062.}}
\lref\Seiberg{N. Seiberg, Nucl. Phys. {\bf B303} (1988) 286.}
\lref\Dubrovin{B. Dubrovin, ``Geometry and integrability of
topological--antitopological fusion", INFN preprint, INFN-8-92-DSF (1992).}
\lref\GPM{B. Greene, D. Morrison, and R. Plesser, in preparation.}
\lref\GPmirror{B. Greene and R. Plesser, Nucl. Phys. {\bf B338} (1990) 15.}
\lref\phases{E. Witten, ``Phases of N=2 Theories in Two Dimensions,"
{\it Nucl. Phys.} {\bf B403} (1993) 159, {\tt hep-th/9301042}.}
\lref\Vafa{C. Vafa, ``String Vacua and Orbifoldized LG models,''
{\it Mod. Phys. Lett.} {\bf A4} (1989) 1169.}
\lref\Ken{K. Intriligator and C. Vafa, ``Landau-Ginzburg Orbifolds,''
{\it Nucl. Phys.} {\bf B339} (1990) 95.}
\lref\Us{S. Kachru and E. Witten, ``Computing The Complete Massless
Spectrum Of A Landau-Ginzburg Orbifold,''
{\it Nucl. Phys.} {\bf B407} (1993) 637, {\tt hep-th/9307038}.}
\lref\WitMin{E. Witten, ``On the Landau-Ginzburg Description of N=2
Minimal Models,'' IAS preprint
IASSNS-HEP-93/10, {\tt hep-th/9304026}.}
\lref\Fre{P. Fr\'e, F. Gliozzi, M. Monteiro, and A. Piras, ``A
Moduli-Dependent Lagrangian For (2,2) Theories On Calabi-Yau n-Folds,''
{\it Class. Quant. Grav.} {\bf 8} (1991) 1455; P. Fr\'e, L. Girardello,
A. Lerda, and P. Soriani, ``Topological First-Order Systems With
Landau-Ginzburg Interactions,'' {\it Nucl. Phys.} {\bf B387} (1992)
333, {\tt hep-th/9204041}.}
\lref\Greene{B.R. Greene, ``Superconformal Compactifications in Weighted
Projective Space,'' {\it Comm. Math. Phys.} {\bf 130} (1990) 335.}
\lref\fendley{P. Fendley and K. Intriligator, ``Central Charges
Without Finite Size Effects,'' Rutgers preprint RU-93-26, {\tt
hep-th/9307101}.}
\lref\OldWit{E. Witten, ``New Issues in Manifolds of SU(3) Holonomy,''
 {\it Nucl. Phys.} {\bf B268} (1986) 79.}
\lref\Pasquinu{S. Cecotti, L. Girardello, and A. Pasquinucci,
``Non-perturbative Aspects and Exact Results for the N=2 Landau-Ginzburg
Models,'' {\it Nucl. Phys.}~{\bf B338} (1989) 701, ``Singularity
Theory and N=2 Supersymmetry,'' {\it Int. J. Mod. Phys.} {\bf A6} (1991)
2427.}
\lref\KT{A. Klemm and S. Theisen, ``Mirror Maps and Instanton Sums for
Complete Intersections in Weighted Projective Space,'' Preprint LMU-TPW
93-08, {\tt hep-th/9304034}. }
\lref\VafaQ{C. Vafa, ``Quantum Symmetries of String Vacua,''
{\it Mod. Phys. Lett.}
{\bf A4} (1989) 1615. }
\lref\HWPmirrors{D. Morrison, ``Picard-Fuchs Equations and Mirror Maps for
Hypersurfaces," In {\it Essays on Mirror Manifolds}, ed. S.--T. Yau,
(Int. Press Co., 1992) {\tt alg-geom/9202026}\semi
A. Font, ``Periods and Duality Symmetries in Calabi-Yau
Compactifications,'' {\it Nucl. Phys.} {\bf B391} (1993) 358, {\tt
hep-th/9203084}\semi
A. Klemm and S. Theisen, ``Considerations of One Modulus Calabi-Yau
Compactification: Picard-Fuchs Equation, K\"ahler Potentials and Mirror
Maps,"
{\it Nucl. Phys.} {\bf B389} (1993) 153, {\tt hep-th/9205041}.}
\lref\flops{P. Aspinwall, B. Greene and D. Morrison, ``Multiple Mirror
Manifolds and Topology Change in String Theory," {\it Phys. Lett.} {\bf 303B}
(1993) 249, {\tt hep-th/9301043}.}
\lref\flopsII{P. Aspinwall, B. Greene and D. Morrison, ``Calabi-Yau Moduli
Space, Mirror Manifolds and Spacetime Topology Change in String Theory,"
IAS and Cornell preprints IASSNS-HEP-93/38, CLNS-93/1236, to appear.}
\lref\cvetic{M. Cvetic, ``Exact Construction of (0,2) Calabi-Yau
Manifolds,'' {\it Phys. Rev. Lett.} {\bf 59} (1987) 2829.}
\lref\Miron{J. Distler, B. Greene, K. Kirklin and P. Miron, ``Calculating
Endomorphism Valued Cohomology: singlet spectrum in superstring models,"
{\it Comm. Math. Phys.} {\bf 122} (1989) 117.}
\lref\miracles{M. Dine and N. Seiberg, ``Are (0,2) Models String Miracles?,"
{\it Nucl. Phys.} {\bf B306} (1988) 137.}
\lref\GrPl{B.R. Greene and M.R. Plesser, ``Mirror Manifolds: A Brief
Review and Progress Report,'' {\tt hep-th/9110014}.}
\lref\masses{P. Candelas, X. De la Ossa and collaborators, to appear.}
\lref\eva{E. Silverstein and E. Witten, ``Global U(1) R-Symmetry and Conformal
Invariance of (0,2) Models," IASSNS-94/4,PUPT-1453, {\tt hep-th/9403054}.}
\lref\WittenElGen{E. Witten, ``Elliptic Genera and Quantum Field
Theory,'' {\it Comm. Math. Phys.} {\bf 109} (1987) 525\semi
``The Index of the Dirac Operator in Loop Space,'' in {\it Elliptic
Curves and Modular Forms in Algebraic Topology}, P.S. Landweber ed.,
Lecture Notes in Mathematics 1326 (Springer-Verlag, 1988).}

\lref\Mohri{T. Kawai and K. Mohri, ``Geometry of (0,2) Landau-Ginzburg
Orbifolds,'' KEK preprint, {\tt hep-th/9402148}.}
\lref\Nemeschansky{D. Nemeschansky and N. Warner, ``Refining the
Elliptic Genus,'' USC preprint, {\tt hep-th/9403047}.}
\lref\Morse{C. Vafa, ``c-theorem and the Topology of 2d QFTs,'' {\it Phys.
Lett.}
{\bf 212B} (1988) 28 \semi
S. Das, G. Mandal, and S. Wadia, ``Stochastic Differential Equations on
Two-Dimensional Theory Space and Morse Theory,'' {\it Mod. Phys. Lett.}
{\bf A4} (1989) 745.}
\lref\BottHomotopy{R. Bott, `` Nondegenerate Critical Manifolds,"
{\it Ann. Math.} {\bf 60} (1954) 248\semi
R. Bott, ``The Stable Homotopy of the Classical Groups,"
{\it Ann. Math.} {\bf 70} (1959) 313.}

\newsec{Introduction}

Conformal field theories with (0,2) worldsheet supersymmetry
are of great interest because of their role in constructing
string-based models of elementary particle physics with spacetime
supersymmetry \Banks.  Calabi-Yau $\sigma$-models with the vacuum gauge
connection identified with the spin connection actually give rise to
(2,2) superconformal field theories \CHSW, and the moduli spaces of such
solutions have been explored in great detail over the past several years
\refs{\StromSpec,\CandMod,\phases}.  Their (0,2) generalizations, which include
Calabi-Yau $\sigma$-models with more general choices for the gauge field
vacuum expectation value \OldWit\ (at least to all orders in
$\sigma$-model perturbation theory), have remained largely mysterious.

In some recent papers, techniques which allow one to study string
compactifications on (0,2) supersymmetric Landau-Ginzburg orbifolds have
been developed and exploited \refs{\phases,\Us,\DK}.  However, although it has
been made plausible that the (0,2) Landau-Ginzburg models studied in
\refs{\phases,\Us,\DK}\ do
indeed represent classical solutions of string theory, no
rigorous proof of their existence as conformal field theories has been
supplied.  Especially in view of the fact that generic (0,2)
Calabi-Yau $\sigma$-models might be destabilized by worldsheet
instantons \DSWW, one would like to have such a proof.

In \S2, we prove that a large class of (0,2) deformations of (2,2)
Landau-Ginzburg orbifolds have nontrivial infrared fixed points.
This is accomplished by showing that they correspond to exactly flat
directions in the spacetime superpotential of the (2,2) theory.
Symmetry considerations analogous to those which
were used in \miracles\ allow us to infer
the existence of many such flat directions.
The novelty is that we make use of the quantum
R-symmetry which is characteristic of Landau-Ginzburg orbifolds \VafaQ.
As an example, we discuss the (0,2) moduli space of the
quintic where, in addition to the 101 complex structure deformations, 200
extra $E_{6}$ singlets are allowed to assume
arbitrary expectation values. This
confirms our intuition
that the space of (2,2) \LG\ orbifolds is but a small subspace of the space of
(0,2) models.

In \S4, we use considerations analogous to those of \S2 in the case of
(0,2) \LG\ theories which are not obviously deformations of (2,2)
theories.  We cannot rigorously prove that such theories exist
as conformal theories by the technique of \S2, since we
cannot choose an expansion point which we know to be conformal.
However, by assuming that one (0,2) \LG\ theory exists as a conformal
theory, we will be able to prove that the neighboring (0,2) theories
(obtained by changing the parameters in the Landau-Ginzburg
superpotential) are also conformally invariant, i.e. that the parameters
in the (0,2) superpotential do indeed correspond to flat directions in the
spacetime superpotential.  This is all discussed in the context of a
particular example studied in detail in \DK, which at large radius
corresponds to a (0,2) theory on a complete intersection Calabi-Yau
manifold in $W{\BP}^{5}_{1,1,1,1,2,2}$.

Having found in \S2\ a large space of exactly marginal (0,2) deformations of
the (2,2)
\LG\ theory, we would like to know to what extent it is possible to turn on
the remaining  $E_6$ singlets. In particular, we would like to know what, if
anything, of this picture persists when we turn on the \Ka\ modulus.
 In \S3, we use the approach of \SeiNat\ to study the couplings of the
twisted sector singlets (including the \Ka\ modulus) in the particular case
of the quintic. We use the $SU(5) \times
SU(5)$ symmetry which arises when one turns off the worldsheet superpotential
to constrain the dependence of the superpotential for the twisted sector
singlets on the untwisted moduli. This allows us to argue both that the 224
$E_{6}$ singlets related to
${\rm H^{1}}(\ET)$
remain massless throughout the \Ka\ moduli space of the
(2,2) quintic, and that the three-point couplings of
these singlets also vanish.   However, the four-point couplings of the 24
twisted sector singlets related to
${\rm H^{1}}(\ET)$ are nonzero, indicating that they are not moduli. Our
arguments are not powerful enough to prove the existence of new exactly flat
directions at arbitrary radius, but we return to that question using other
techniques in \S5.

After providing strong evidence that the (0,2) Landau-Ginzburg models do
indeed correspond to bona-fide CFTs in \S2 and \S4, we turn in \S5 to
the subject of (0,2) models at finite radius.  Making some very plausible
assumptions about the renormalization group flow of the (0,2)
linear
$\sigma$-models we have been studying,
and restricting ourselves
to models with 1 (complex) dimensional \Ka\ moduli
spaces, we are able to prove that
the large (0,2) moduli spaces we have found at the \LG\ radius
exist at all values of the \Ka\ modulus.
For example, the full 102-dimensional (2,2) moduli space of the quintic is
actually a submanifold of a 302 dimensional (0,2) moduli space.
Our proof uses elementary ideas of Morse theory, applied to
Zamolodchikov's c-function.\foot{Morse theory and the c-function have
also come together elsewhere in the string literature \Morse.}

In \S6, we
discuss some interesting questions which remain to be
answered by future explorations of (0,2) moduli space.

\newsec{(0,2) Deformations}

R-symmetries are particularly useful tools in establishing the
existence of exactly flat directions in supersymmetric field theories.
Consider a supersymmetric gauge theory with some collection of
chiral fields
$\Phi_{1},\dots,\Phi_{M}$ and suppose furthermore that this theory has
an R-symmetry
under which the superfields
$\Phi_{1},\dots,\Phi_{N}$ ($N\leq M$) are invariant.
Then since the spacetime
superpotential ${\cal W}$ is not invariant under the symmetry, no terms
of the form $f(\Phi_{1},\dots,\Phi_{N})$ can appear in the
superpotential.
If furthermore no terms of the form $f(\Phi_{1},\dots,\Phi_{N})X$
can appear in the spacetime superpotential for
$X=\Phi_{N+1},\dots,\Phi_{M}$, then the
directions in field space corresponding to the scalar components of
$\Phi_{1},\dots,\Phi_{N}$ are necessarily F-flat.
Therefore,
as long as the constraints of D-flatness are also satisfied,
the VEVs of the scalar components of
$\Phi_{1},\dots,\Phi_{N}$ parametrize a space of degenerate ground
states for the supersymmetric theory in question.

In conformal perturbation theory, one needs to examine certain correlation
functions of the zero-momentum vertex operators corresponding to the fields
$\Phi_i$ in order to show that they correspond to exactly marginal
deformation of the worldsheet superconformal field theory ({\it i.e.}~that
they preserve superconformal invariance). But this is exactly the same
condition as demanding the {\it spacetime} equations of motion be satisfied.
Here is where spacetime supersymmetry comes to our aid. We only need to check
F- and D-flatness to assure ourselves that the spacetime equations of motion
are satisfied, and this involves examining a much smaller and more tractable
set of worldsheet correlation functions than would be the case without
spacetime supersymmetry. In fact, if the fields $\Phi_i$ are gauge-singlets,
D-flatness is automatic, so we only need to check F-flatness.

It is well known that many (2,2) Calabi-Yau compactifications possess,
at special points in their complex structure moduli space ${\cal
M}_{C}$, extra
``classical'' discrete R-symmetries (which are essentially symmetries of
the defining equations of the Calabi-Yau in some ambient
weighted projective space) \GSW.  In the context of string
compactification on the quintic, these R-symmetries have been used
to prove the existence of exactly flat (0,2) directions at
special points in ${\CM}_{C}$.
In particular, since conformal perturbation theory about an
interior point in (2,2) moduli space does not miss any non-perturbative
$\sigma$-model effects, demonstrating the existence of a
flat (0,2) direction by such macroscopic reasoning $\it guarantees$ that
the corresponding (0,2) theories are not destabilized by worldsheet
instantons \miracles.

The Landau-Ginzburg orbifolds are distinguished by the fact that they
$\it all$ possess at least one discrete R-symmetry, namely the
``quantum'' symmetry which counts the twisted sector $k$ of origin of
the various physical states \VafaQ.
The VEVs of the massless gauge-singlet fields which arise in
the untwisted sector and are uncharged under the quantum
symmetry are therefore guaranteed to be moduli of the spacetime
supersymmetric field theory, and on the string worldsheet these fields will be
represented by mutually integrable moduli of the conformal field
theory.  The massless singlets which arise in the untwisted
sector of the
(2,2) \LG\ theories typically include many $E_{6}$ singlet fields which
are related to neither complex structure nor \Ka\ structure
deformations -- at
large radius,
these modes are related to the cohomology group ${\rm H^{1}}({\ET})$.
Giving VEVs to these fields breaks the (2,2) worldsheet
supersymmetry to (0,2) supersymmetry.

For concreteness, let us focus attention on the quintic hypersurface in
$\cp{4}$. The \LG\ theory is a point of enhanced symmetry in the \Ka\ moduli
space. One normally says that the \LG\ orbifold has a $\BZ_5$
quantum symmetry, but
since one needs to include both NS and R sectors for the
left-movers, there are actually 10 sectors in the \LG\ orbifold. So one might
better think of the quantum symmetry as
$\BZ_{10}=\BZ_2\sdp\BZ_5$. Actually, this definition of the quantum symmetry
is a little awkward because the different components, under the decomposition
$E_6\supset SO(10)\times U(1)$, of a given $E_6$ representation transform
with different weight under this $\BZ_{10}$ symmetry. To fix this, we can
compose this symmetry with an element of the center of $E_6$, to obtain a
$\BZ_{30}=\BZ_3\sdp\BZ_{10}$ symmetry which acts homogeneously on $E_6$
multiplets. In the language of \Us, this $\BZ_{30}$ is generated by
\eqn\eSqdef{S_Q=\ex{2\pi i(3 k-2 \ql)/30}}
where $k=0,\dots,9$ labels the sector number of the \LG\ orbifold, and $\ql$ is
the left-moving $U(1)$ charge.\foot{The states found in
\Us\ were the massless spacetime fermions;  for a fermion which comes
from
the $(k+1)$st twisted sector, its scalar superpartner comes from the $k$th
twisted sector.}
The charges of the various massless
multiplets  under $S_Q$ are listed in Table 1 as integers $\in \BZ/30\BZ$.

\def\tablerule{\omit&\multispan{12}{\tabskip=0pt\hrulefill}&\cr}
$$\vbox{\offinterlineskip\tabskip=0pt\halign{\hskip 1.0in
$#$\quad&\vrule #&\quad\hfil $\strut #$\hfil\quad &\vrule #&
\quad\hfil $#$\hfil\quad
&\vrule # &\quad\hfil $#$\hfil\quad &\vrule #&\quad\hfil $#$\hfil\quad &\vrule
#&\quad\hfil
$#$\hfil\quad &\vrule{} \vrule #&\quad\hfil $#$\hfil\quad &\vrule #\cr
&\omit&{\bf 27}&\omit&\overline{\bf 27}
&\omit&C,S&\omit&R&\omit&S'&\omit&\CW&\omit\cr
\tablerule
S_Q&&-2&&8&&0&&6&&6&&-6&\cr
\tablerule
\noalign{\bigskip}
\noalign{\narrower\noindent{\bf Table 1:} Charges ($\in \BZ/30\BZ$) of the
spacetime matter multiplets, and of the spacetime
superpotential, $\CW$, under $S_Q$, the ``quantum"
R-symmetry present at the \LG\ point.
}
 }}$$

On the world sheet, $S_Q$ simply enforces the fact that sector number is
conserved mod 10 in correlation functions. It is easy to see that, in {\it
spacetime},
$S_Q$ generates a discrete R-symmetry, under which the spacetime
superpotential has charge $-6$ mod 30. That is, one should add 3 to the
entries in Table 1 to obtain the $S_Q$-charge of the corresponding
right-handed fermions in these chiral multiplets. Clearly ${\bf 27}^3$ and
$\overline{\bf 27}^3$ are couplings in the superpotential allowed by the
discrete
$R$ symmetry, whereas, say, ${\bf 27}^2\overline{\bf 27}^2$ is not.

In Table 1, we have divided the 224 singlets corresponding to elements of
${\rm H^{1}}({\ET})$ into the 200, denoted by $S$, which arise in the untwisted
sector of the
\LG\ orbifold, and the 24, denoted by $S'$, which arise in the $k=2$
twisted sector.

 Let us recall how this distinction arises \Us. The
$\H{1}{\ET}$ singlets can be identified with operators of the form
\eqn\eSdef{S=\lambda^i P_i(\phi)}
 where
$P_i(\phi)$ are a set of five quartics satisfying
\eqn\ePrest{\phi^i P_i(\phi)=0~.}
There are $5\times70-126=224$ operators \eSdef\ satisfying \ePrest. However,
in the \LG\ theory, precisely 24 of these are $\bar Q_+$-trivial. Namely, we
need to mod
out the polynomials satisfying \ePrest\ by the equivalence relation
($W(\phi)$ is the (2,2) worldsheet superpotential)
\eqn\ePequiv{P_i(\phi)\sim P_i(\phi)+A_i {}^j\partial_j W
-{1\over5}\partial_i(\phi^j A_j {}^k\partial_k W)}
for an arbitrary traceless matrix $A_i{}^j$.

The 24 singlets from the twisted sector which ``replace" the missing singlets
from the untwisted sector take the form
\eqn\eSprimedef{S'_{i j}=
(\bar\lambda^i_{-3/10}\phi^j_{-1/5}-{\textstyle{1\over5}}\delta^{i
j}\bar\lambda_{-3/10}\cdot\phi_{-1/5})\CS_2}
where $\CS_2$ is the field that creates the ground state of the twisted sector.

Let us now see what restrictions on the $S$-dependence of the spacetime
superpotential are imposed by this quantum R-symmetry. First of all, since
both the $C$s and the $S$s are neutral under $S_Q$, and a term in $\CW$ must
have charge $-6$ mod 30, we see that, at the \LG\ point, no term in the
superpotential of the form $\CW=f(C,S) X+\dots$, where $X$ is any singlet, is
allowed. Thus, at the \LG\ point, for an {\it arbitrary} complex structure,
all 200 $S$s
correspond to flat directions in the superpotential. In the
conformal field theory, these are exactly marginal operators, which break
(2,2) superconformal symmetry to (0,2), while preserving $E_6$ as the
spacetime gauge group.

We should note at the same time that
$\CW={\bf 27}\overline{\bf 27} S^n+\dots$ is also forbidden, so that turning on
$S$ does not cause the $\overline{\bf 27}$ and a {\bf 27} to pair up and get a
mass.

At the \LG\ point in its \Ka\ moduli
space, we have found that the quintic has, at least, a 301 dimensional (0,2)
moduli space of $E_6$ preserving (0,2) deformations. Two questions naturally
arise:
\item{$\bullet$}Do these (0,2) deformations remain exactly marginal
when we turn on the \Ka\ modulus?
\item{$\bullet$}Are any of the 24 singlets, $S'$, which occur in the same
twisted sector as the \Ka\ modulus, mutually integrable with the deformations
we have found?

We will address these questions in the next section. It will turn out that the
answer to the second question is {\it no}. The $S'$ are charged under the
quantum R-symmetry, so it is possible for the to have a nontrivial
superpotential,  spoiling their flatness. The quantum R-symmetry dictates that
the lowest possible term is quartic, $\CW=S'^4+\dots$

The answer to the first question is
{\it likely, yes}, but we will not see that until \S5.
What would it mean if the answer to the first question turned out to be
{\it no}? It would imply that the \LG\ theory is a multicritical ``point"
in the (0,2) moduli space ${\cal M}_{(0,2)}$.  ${\cal M}_{(0,2)}$ would consist
of two components -- a 102-dimensional space of (2,2) symmetric theories, and
a 301-dimensional space of what are generically (0,2) symmetric theories --
which meet along the 101-dimensional locus of (2,2) symmetric \LG\ theories.
This is schematically depicted in \tfig\figone.

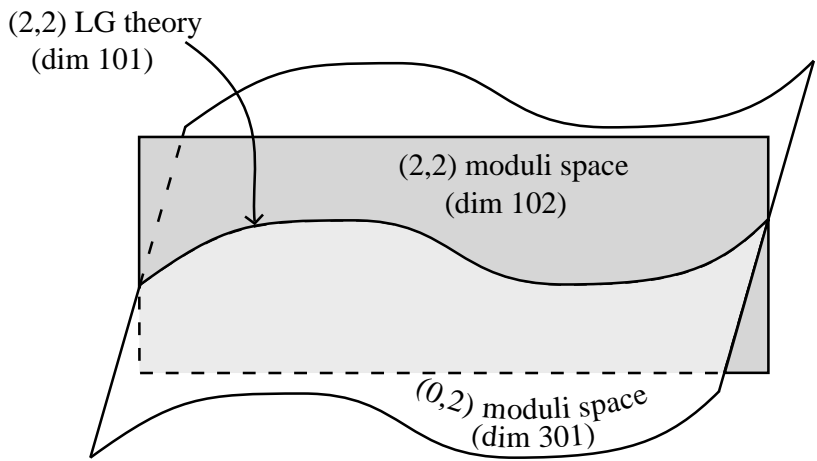
\ifigure\figone{Schematic picture of the moduli space of the quintic,
showing the intersection of $\CM_{(2,2)}$ and ${\cal
M}_{(0,2)}$ along the locus of (2,2) \LG\ theories.}{modspace.eps}{2.5}

One might ask what this picture looks like under mirror symmetry. The \LG\
locus in the \Ka\ moduli space is simply the locus $\psi=0$ in the complex
structure moduli space of the mirror quintic, where the polynomial is of
Fermat form. The 200 singlets in question all arise in blowing up the
singularities of the mirror, as do 100 of the 101 \Ka\ moduli. All of these
are mutually-integrable marginal perturbations. So we see that, for the
Fermat form of the quintic mirror, there is a 200 parameter family of (0,2)
deformations at {\it arbitrary} radius!

One intriguing possibility -- not realized in this example -- is that a
generic (0,2) compactification might freeze the radius at some Planckian
value. This is what would seem to happen when the theory is formulated on the
original quintic, if
the answer to
the first question was no: The theory would seem
to be stuck at the \LG\ radius. However,
in
the mirror picture, it is clear that there are still some directions in
$\CM_{(0,2)}$ which ought to be interpreted as ``large radius".

\newsec{Twisted Sector Singlet Couplings}

Consider a general point in the 301 dimensional moduli space of (0,2) \LG\
theories that we have found on the quintic. The worldsheet superpotential can
be written in (0,2) superspace as
$$\int d^2zd\theta\ \Lambda^a F_a(\Phi)=\int d^2zd\theta\
F_{aijkl}\Lambda^a\Phi^i\Phi^j\Phi^k\Phi^l$$
If  we neglect the superpotential, the theory possesses an $SU(5)\times
SU(5)$ symmetry under which the $\Lambda$s and the $\Phi$s rotate
independently.

Properly speaking, we should also include a $U(1)\times
U(1)$ phase symmetry as well. One of these $U(1)$s is generated by our
old friend $\ql$, and doesn't teach us anything new.
Unfortunately, the peculiar quantization \WitMin\ of the zero mode of the
scalar
field (which is noncompact when we turn off the $F$s) spoils the conservation
of the remaining U(1) (even when we neglect the explicit breaking by the
$F$s). So, in the end, the only new symmetry we have to exploit is
$SU(5)\times SU(5)$.

The coupling constants
$F_{aijkl}$  break this symmetry explicitly, transforming as the
$(\overline{\bf 5}, {\bf 70}')$ representation. Since this is the {\it only}
source of $SU(5)\times SU(5)$-breaking in the theory,
we will be able to constrain the dependence on the $F$s of various correlation
functions
of the massless multiplets by demanding that they transform correctly under
$SU(5)\times SU(5)$.

We will be particularly interested in the couplings of the 25 singlets from the
twisted sector. We denote them collectively by
$$S'^{\bar ai}=\bar\lambda^a\phi^i\CS_2$$
and clearly they transform as the $(\overline{\bf 5},{\bf 5})$ representation.

We saw in the previous section that the quantum symmetry dictates that the
lowest nonvanishing term in the superpotential for the $S'$s is at least
quartic. So we will be interested in computing an $\langle S'^4\rangle$
coupling.

Before we launch in, however, there is clearly a subtlety we must deal with.
The spacetime superpotential is a {\it section of a line bundle} over the
moduli space (that is, over the space of $F$s). We therefore need to supply
some trivialization of that line bundle in order to specify it.
There is no obvious candidate for such a trivialization, even locally on the
space of the $F$s.

To evade this ambiguity, we will simply note that {\it ratios} of
superpotential
couplings transform as sections of a {\it trivial} line bundle, and so are
canonically-defined (up to scale) as functions on the moduli space.
Since we are interested in exploiting the $SU(5)\times SU(5)$ transformation
properties, a natural candidate to normalize our correlation functions
is the $\langle\overline{\bf 27}^3\rangle$ coupling\foot{The somewhat
attentive reader might wonder how the statements of this paragraph are to be
reconciled with the oft-repeated statement that ``the $\overline{\bf 27}^3$
coupling
is independent of the complex structure moduli" \Exact. In the notation of
\Trieste, the spacetime superpotential is a section of the  line bundle
$\tilde L^3$. The $\overline{\bf 27}$s are sections of $T_{\CM_R}\otimes L$,
the tangent bundle to the \Ka\ moduli space, twisted by the line bundle $L$.
So the cubic {\it coupling} is a linear map from $S^3(T_{\CM_R}\otimes L)\to
\tilde L^3$. In other words, it is a section of $S^3(T^*_{\CM_R})\otimes
(L^{-1}\otimes \tilde L)^3$. But, restricted to the complex structure moduli
space, $T_{\CM_R}$ and $L^{-1}\otimes \tilde L$ are trivial bundles. Hence it
makes
sense to say that the coupling, a section of a trivial bundle, is a constant.
 }, which is, after all, an
$SU(5)\times SU(5)$ singlet.
\eqnn\eratio\
So we will denote\foot{A related point
is that we need to pin down the ambiguity in the $F$-dependence of the
normalization of the operator $\CS_2$ (and the corresponding fermion state
$\ket{3}$) which go into defining the vertex operators for the $S'$s. But the
$\overline{27}^3$ coupling can be represented, for instance, as the matrix
element $\melt{3}{\CS_4}{3}$. So if we define the relative normalization
to be such that $\CS_4$ appears with unit coefficient in the OPE of two
$\CS_2$s
then all ambiguity in the normalization of these operators disappears from the
ratio \eratio.}
$$\eqalignno{\langle\langle S'^4\rangle\rangle
&={\langle S'^4\rangle\over\langle
\overline{\bf 27}^3\rangle}&\eratio\cr}$$

Our task is to determine the dependence of this correlation function on the
$F$s. Clearly, $SU(5)\times SU(5)$ symmetry is not going to be enough
to determine the complete dependence for us. It is possible to write a
polynomial in the $F$s (the lowest degree of such a polynomial is 10) which
is an $SU(5)\times SU(5)$ singlet. Our determination of the correlation
function
is therefore ambiguous up to multiplication by an arbitrary $SU(5)\times SU(5)$
 singlet {\it function} of the $F$s.

Modulo this ambiguity, we can still place some powerful constraints on the
$SU(5)\times SU(5)$ nonsinglet part of the $F$ dependence.

\item{1)} Analyticity. The spacetime superpotential depends holomorphically
on the moduli, and so is a function of the $F$s, but not the $\bar F$s.
\item{2)}Quintality.
The correlation function \eratio\ transforms as the $4^{th}$ symmetric power
of $(\overline{\bf 5},{\bf 5})$. By quintality,
a polynomial in the $F$s (which, recall, are in the
$(\overline{\bf 5}, {\bf 70}')$ representation) which transforms in this
representation must have degree $4+5n$.

\item{3)} Flatness of the $(2,2)$ moduli. Recall that the \Ka\ modulus
$R=\delta_{\bar a i}S'^{\bar a i}$. Under the diagonal
$SU(5)\subset SU(5)\times SU(5)$,
the $F_{aijkl}$ transform as ${\bf126}'\oplus {\bf224}$. The (2,2) theory
is obtained by setting to zero the ${\bf224}$ piece, and considering
$F_{aijkl}$
which are totally symmetric on their five indices. Since $R$ is indeed a
modulus of the $(2,2)$ theory, $\vev{\vev{R^4}}=\delta_{\bar a_1 i_1}\dots
\delta_{\bar a_4 i_4}\vev{\vev{S'^{\bar a_1 i_1}\dots S'^{\bar a_4i_4}}}$
must vanish when we set the $F$ to their (2,2) symmetric values.

\item{4)} Flatness of ``twisted $(2,2)$" moduli. The above case corresponded
to choosing polynomials
 \eqn\esquare{F_a(\phi)=\partial_a W(\phi)~.}
However, the most general form
for the $F$s which preserves a (2,2) supersymmetry is
 \eqn\etest{F_a(\phi)= ((U^T)^{-1})_{a}{}^i\partial_i W(V^{-1}\phi) }
for some invertible matrices $U,V$. When $U=V$, this is just a harmless
$GL(5)$ transform of \esquare, and the left-moving supersymmetry is the
standard one. When
$U^{-1}=V^5$, this is the global form of the deformations in ${\rm H^{1}}
(\ET)$ in
the ideal \ePequiv. Expanding $U=\Bid- A^T$, $V=\Bid +{1\over 5}A^T$, we find
that
$F_a(\phi)$ is a deformation of $\partial_a W(\phi)$ by an element of the
ideal \ePequiv.  This is a
$\bar Q_+$-trivial deformation, but it forces us to {\it redefine} the
left-moving supersymmetries. Instead of
$G^+=
\delta_{\bar a i}\left(-{4\over 5}\bar\lambda^a\partial\phi^i+{1\over
5}\partial\bar\lambda^a\phi^i\right)$
we have
\eqn\etwisted{G^+= \delta_{\bar a
a}(UV^{-1})^a {}_i\left(-{4\over 5}\bar\lambda^a\partial\phi^i+{1\over
5}\partial\bar\lambda^a\phi^i\right)~.}
Naturally, then, we should redefine the \Ka\ modulus to be
\eqn\enewR{R=\delta_{\bar a
a}(UV^{-1})^a {}_i\bar\lambda^a\phi^i\CS_2}
Contracting the $S'^4$ correlation function with $\delta_{\bar a
_1a_1}(UV^{-1})^{a_1}{}_{i_1}\dots$, we must find that the redefined
$R^4$
coupling  vanishes for $F$s of the form \etest.

\item{5)} In addition to the $R^4$ coupling,
the $\vev{\vev{S'^{\bar a i}R^3}}$ coupling must
also vanish. This is clear, since we know that the
deformed theory is conformal, so the 1-point functions (in this case, of
$S'$) vanish.

\item{6)}The correlation function \eratio\ should transform covariantly under
the $G= \left(GL(5)\times GL(5)\right)/\BC^*$ action \etest. It is convenient
to fix the
$\BC^*$ symmetry by choosing the gauge
$$\det(U^{-1}V)=1\quad .$$
Then, under the $G$ action,
\eqn\etransform{\eqalign{
\vev{\vev{S'^{\bar a_1 i_1}\dots S'^{\bar a_4 i_4}}}\to
&
(\delta^{\bar a_1 a_1}((U^T)^{-1})_{a_1}{}^{b_1}\delta_{\bar b_1 b_1})
\dots
(\delta^{\bar a_4 a_4} ((U^T)^{-1})_{a_4}{}^{b_4}\delta_{\bar b_4
b_4})\times\cr
&\times(V)^{i_1}{}_{j_1}\dots
(V)^{i_4}{}_{j_4}
\vev{\vev{S'^{\bar b_1 j_1}\dots S'^{\bar b_4 j_4}}}
\cr }}

Together, these are quite stringent constraints. Up to an $SU(5)\times
SU(5)$ singlet function of the $F$s, $f(F)$, we obtain
\eqn\efourpt{\eqalign{\vev{\vev{S'^{\bar a_1 i_1}S'^{\bar a_2 i_2}S'^{\bar a_3
i_3}S'^{\bar a_4 i_4}}}=&
\delta^{\bar a_1 a_1}\delta^{\bar a_2 a_2}
\delta^{\bar a_3 a_3}\delta^{\bar a_4 a_4}\times\cr
&\times\epsilon^{i_1 j_1 k_1 l_1 m_1}\epsilon^{i_2 j_2 k_2 l_2 m_2}
\epsilon^{i_3 j_3 k_3 l_3 m_3}\epsilon^{i_4 j_4 k_4 l_4 m_4}\times\cr
&\times F_{a_1 j_1 j_2 j_3 j_4}F_{a_2 k_1 k_2 k_3 k_4}
F_{a_3 l_1 l_2 l_3 l_4}F_{a_4 m_1 m_2 m_3 m_4}\ f(F)\cr}}

{}From \etransform, we learn that, under the $G$ action,
$f(F)\to \det(V)^{4}f(F)$.  An example of an invariant
function which transforms with this weight is
$$\eqalign{f(F)=\bigl(
\epsilon^{a_1\dots a_5}\epsilon^{a_6\dots a_{10}}&\epsilon^{i_2\dots i_6}
\epsilon^{i_7\dots i_{10}i_1}\dots\epsilon^{l_5\dots l_9}\epsilon^{l_{10}l_1
\dots l_4}\times\cr
&\times F_{a_1i_1j_1k_1l_1}\dots F_{a_{10}i_{10}j_{10}k_{10}l_{10}}
\bigr)^{-2/5}
}$$

More generally, we can consider a $(4+5n)$-point function,
\eqn\ehigherpt{\vev{\vev{S'^{\bar a_1 i_1}S'^{\bar a_2 i_2}S'^{\bar a_3
i_3}S'^{\bar a_4 i_4} R^{5 n}}}}
Note that the \Ka\ modulus $R$, being the tangent
vector to the \Ka\ moduli space, can be normalized in a fashion independent
of the complex structure. That is, the corresponding (0)-picture
vertex operator can be defined to be {\it independent} of the $F$s. All of
the conditions 1)--6) generalize to \ehigherpt.
Thus, up to, perhaps, a different choice of the function $f(F)$, \ehigherpt\
must have exactly the same form as the right hand side of \efourpt.

Certainly, we haven't {\it proven} that \efourpt\ is the only $SU(5)\times
SU(5)$ structure that can occur. However, we have not been able to find a
structure at higher orders which satisfied conditions 1)-6) and could not be
reduced to the form \efourpt. Perhaps, at sufficiently high order in the
$F$s, one exists, but we have not been able to find it.

In any case, the striking feature  of \efourpt\ is that, not only does it
vanish when three or four of the $S'$s are in fact $R$s, but it vanishes
when {\it any } of the $S'$s are $R$s. Thus the zero, one, {\it two and three}
point functions of the $S'$s vanish at arbitrary values of the \Ka\ modulus!
Even if \efourpt\ turns out not to be unique and there exist higher order
invariants not reducible to  \efourpt, it is very likely that those invariants
share this property.

One can go further than this. Differentiating \efourpt\ with respect to the
$F$s, and then setting them equal to their (2,2)-symmetric values
 has the effect of inserting untwisted sector singlets ($S$s, or $C$s) into
the correlation function. Schematically,
$$\partial_F\vev{\vev{S'^{4} R^{5n} }}=\vev{\vev{S S'^{4} R^{5n} }}
-\vev{\vev{S'^{4} R^{5n} }}\vev{\vev{S \overline{\bf 27}^3 }}
 $$
We can contract indices appropriately to turn some of these $S'$s into $R$s.
Having shown that the zero, one, two and three
point functions of the $S'$s vanish, we see, from explicitly differentiating
the RHS of \efourpt, that the zero, one, two and three
point functions of any {\it combination} of $S'$s and $S$s also vanish.
In particular, we have 224 massless $\H{1}{{\rm End}(T)}$
singlets at an arbitrary point in the
\Ka\ moduli space.

The lowest
nonvanishing singlet couplings in the (2,2) \LG\ theory, consistent with both
the quantum symmetry and
\efourpt\ are: $S^4R^{4+5n}$, $S^3 S'R^{3+5n}$, $S^2S'^2R^{2+5n}$, $S
S'^3R^{1+5n}$,  and
$S'^4R^{5n}$. At finite $R$, these are all quartic couplings among the
$\rm{H^{1}}(\ET)$ singlets,
there being no invariant distinction between the $S$s
and $S'$s at finite $R$. So the general statement is that the superpotential
for the $\rm{H^{1}}(\ET)$ singlets starts at quartic order at an arbitrary
point in the \Ka\ moduli space of the (2,2) theory.

We saw in the previous section that there is, at least, a 200-dimensional
space of $E_6$-preserving (0,2) deformations of the quintic \LG\ theory.
The four point function \efourpt\ is the obstruction to extending this
further to include the 24 singlets in the twisted sector.
It is not a terribly difficult calculation
to compute this obstruction explicitly in, say, the Gepner model.

The correlation function one wants to evaluate is $\langle S'^{\bar1 5}
S'^{\bar2 5}S'^{\bar3 5}S'^{\bar4 5}\rangle$. The $(-1)$-picture vertex
operator for $S'^{\bar1 5}$ is given in the tensor product of minimal models
by the operator
$\mm{0}{-2}{-2}{0}{0}\mm{1}{-1}{0}{1}{0}^3\mm{2}{0}{0}{2}{0}$, and similarly
for the other $S'^{\ib 5}$. We need to shift two of these operators by
$\mm{0}{0}{0}{1}{1}^{5}$ to produce the corresponding $(-1/2)$-picture fermion
vertex operators, and we need to shift one of the remaining vertex operators
by $\mm{0}{0}{0}{0}{0}^4\mm{0}{0}{0}{0}{2}$ to produce the $(0)$-picture
vertex operator which is to be integrated over the worldsheet. Even without
explicitly calculating the integrated 4-point function, we can readily see that
the N=2 minimal model fusion rules are compatible with its being
nonzero.

The hard question, which we have not been able to address, is how many (0,2)
directions are obstructed at finite $R$? At the \LG\ point, it is the 24 $S'$
that are obstructed, whereas the 200 $S$s are unobstructed. When we turn on the
\Ka\ modulus, the distinction between the $S$s and $S'$s is effaced, and some
combinations of these 224 singlets are obstructed. Perhaps all of them are, if
not by this term, then by higher terms in the superpotential which we have not
yet considered.  We will see in \S5\ that this is very likely not the case, and
that, in fact, 200 of them remain unobstructed. However, it is clear
from the
computations of this section that precisely {\it which} combinations
of singlets
are unobstructed is a complicated function of both the
\Ka\ and complex structure moduli.

\newsec{More General (0,2) Theories}

We have seen that the quantum R-symmetry of (2,2) Landau-Ginzburg models is
enough to guarantee that, in many cases, the (2,2) moduli space is a
small part of a much larger (0,2) moduli space.
What can $S_{Q}$ do for us in the context of (0,2) \LG\ theories that
are not obviously obtained as deformations of (2,2) theories \DK?

The \LG\ theories discussed in \DK\ were described by a (0,2) worldsheet
superpotential
of the form
\eqn\esuperpot{\int d^2 z d\theta\ \Sigma^j W_j(\Phi)+\Lambda^a F_a(\Phi)}
defined in such a way that the theory possesses both a U(1) symmetry with
charge $\ql$, and a U(1) R-symmetry with charge $\qr$, where the charges
of the fields are given in Table 2.

\def\tablerule{\omit&\multispan{4}{\tabskip=0pt\hrulefill}&\cr}
$$\vbox{\offinterlineskip\tabskip=0pt\halign{\hskip 1.0in
$#$\quad&\vrule #&\quad\hfil $\strut #$\hfil\quad &
\vrule #&\quad\hfil $#$\hfil\quad &\vrule #\cr
&\omit&\ql&\omit&\qr&\omit\cr
\tablerule
\Phi_i&&q_i&&q_i&\cr
\tablerule
\Lambda^a&&q_a-1&&q_a&\cr
\tablerule
\Sigma^j&&q_j-1&&q_j&\cr
\tablerule
\noalign{\bigskip}
\noalign{\narrower\noindent{\bf Table 2:} $U(1)$ and $U(1)_R$ Charges
of the (0,2) \LG\ superfields.
}
 }}$$
\noindent
The charges are constrained to satisfy \DK
\eqn\echconst{\eqalign{\sum_{j=1}^n(q_j-1)&=-\sum_{i=1}^{n+D+1}q_i\cr
\sum_{a=1}^{r+1}q_a&=1\cr
\sum(q_j-1)^2+\sum q_a^2&=1+\sum q_i^2\cr}}
where $D=3$ for Calabi-Yau threefolds.
In the infrared, the U(1) current $\bar J$, associated to $\qr$, becomes the
U(1) current in the (0,2) superconformal algebra. We can read off the infrared
central charge from the $\bar J$-$\bar J$ anomaly \DK.
The constraints \echconst\
ensure that this gives
\eqn\ecbar{\bar c/3=\sum(q_i-1)^2-\sum q_a^2-\sum q_j^2=D\quad .}
Similarly, $J$, the  U(1) current associated to $\ql$, generates a left-moving
U(1) current algebra in the infrared, whose central extension can be read
off from the $J$-$J$ anomaly:
$$r=\sum(q_j-1)^2+\sum(q_a-1)^2-\sum q_i^2\quad .$$
And, of course, consistency requires that the $J$-$\bar J$ anomaly vanish:
$$0=\sum (q_i-1)q_i-\sum q_j(q_j-1)-\sum q_a(q_a-1)$$
which, again, is assured by \echconst.

In fact, the situation is even better than that \refs{\WitMin,\Us,\DK}.
Even in the off-criticality
theory, the operators
$$\eqalign{T'(z)&=-\sum_i\left(\partial\phi_i\partial\bar\phi_i
+{q_i\over2}
\partial(\phi_i\partial\bar\phi_i)\right)
+\sum_a\left(\lambda_a\partial\bar\lambda_a
-{1-q_a\over2}\partial(\lambda_a\bar\lambda_a)\right)\cr
&\hskip1in+\sum_j\left(\sigma_j\partial\bar\sigma_j
-{1-q_j\over2}\partial(\sigma_j\bar\sigma_j)\right),\cr
J'(z)&=
-\sum_iq_i\phi_i\partial\bar\phi_i+\sum_a(1-q_a)\lambda_a\bar\lambda_a
+\sum_j(1-q_j)\sigma_j\bar\sigma_j}$$
commute with the right-moving supersymmetry generator $\bar Q_+$, and
generate a Virasoro$\sdp\widehat{U(1)}$ algebra on the $\bar Q_+$-cohomology
\foot{That this algebra is satisfied on the quantum level {\it requires} that
the conditions \echconst\ hold \eva.}.
This algebra coincides with the left-moving chiral algebra
in the infrared, and one can again compute, using the free field methods of
\WitMin,
 the Virasoro central charge
$(c=6+r)$, and the $\widehat{U(1)}$ central charge $(r)$, in agreement with
above.

Finally, to embed this theory in a heterotic string theory,
we need to orbifold
the \LG\ theory \esuperpot\ by the $\BZ_{2m}$ group generated by
$\ex{-i\pi \ql}\times (-1)^{F_f}$, where $F_f$ is the fermion number for a set
of $16-2r$ free left-moving Majorana-Weyl fermions which represent the gauge
degrees of freedom
\Us. Here $m$ is the least common denominator of all the charges in Table 2.

But all of these calculations {\it assume} that
(0,2) supersymmetry is unbroken in the infrared limit. If (0,2) supersymmetry
is
spontaneously broken in the infrared, then all bets are off. Since we
don't have as firm a grasp of the dynamics of the theory
\esuperpot\ as we do of the more familiar (2,2) \LG\ theories, any consistency
checks that we can apply should bolster our confidence that (0,2) supersymmetry
is indeed unbroken in the infrared, and that the \LG\ orbifold is a
bona-fide string vacuum.

One such consistency check is provided by the quantum R-symmetry.
Assume that one point
in the moduli space of the (0,2)
\LG\ theory does exist as a (0,2) conformal theory.
As in the (2,2) case discussed in \S2, this (0,2) \LG\ orbifold
will have a quantum
R-symmetry generated on the worldsheet by
\eqn\eSQgen{S_Q=\ex{2\pi i(k r -2\ql)/2 m r}}
where $k=0,1,\dots 2m-1$ is the sector number, and $r$ is the ``rank of the
vacuum gauge bundle" -- $r=3,4,5$ for spacetime
gauge groups $E_6,~SO(10),$ and $SU(5)$.
One finds, as in the (2,2) case, that the R-symmetry guarantees
that all of the singlets
corresponding to $\bar Q_+$-nontrivial deformations of the (0,2)
superpotential \esuperpot\
are indeed flat directions. They all are neutral under the
quantum R-symmetry because they all come from the untwisted sector of the \LG\
orbifold.
So assuming that one point in the moduli space of the (0,2) \LG\ theory
exists, we are able to prove that all of the $\bar Q_{+}$ non-trivial
parameters in the worldsheet superpotential correspond to moduli of the
theory.  This is a non-trivial self-consistency check on the assumption
that these (0,2) \LG\ theories have infrared fixed points with the
desired properties.

For example, consider
the model discussed in detail in \S4.1 of \DK\ (the $Y_{W5;4,4}$
model listed in \Greene).   The Calabi-Yau $\sigma$-model description of
this theory consists of a rank 4 vacuum gauge bundle over a complete
intersection manifold in $W{\BP}^{5}_{1,1,1,1,2,2}$ defined by the
vanishing loci of two degree four polynomials.  Therefore, this theory
yields an $SO(10)$ observable gauge group in spacetime.  At the
Landau-Ginzburg point in its \Ka\ moduli space, it has a $\BZ_{10}$
quantum symmetry (like the quintic).  However, this definition of the
quantum symmetry is somewhat awkward because the different components of
a given $SO(10)$ representation, under the decomposition $SO(10) \supset
SO(8)\times U(1)$, transform with different weight under the $\BZ_{10}$
symmetry.  We can fix this as we did in the case of the quintic, by
multiplying by an element of the center of $SO(10)$, to obtain a
$\BZ_{20}$ symmetry which acts homogeneously on $SO(10)$ multiplets.
This $\BZ_{20}$ is generated by
\eqn\quantwo{S_{Q} = e^{2\pi i ({{2 k-\ql}\over 20})}}
where $k=0, \dots, 9$ labels the sector number of the \LG\ orbifold and
$\ql$ is the left-moving $U(1)$ charge.  The charges of the various
multiplets under $S_{Q}$ are listed in Table 3 as integers
$\in \BZ/20\BZ$.

\def\tablerule{\omit&\multispan{12}{\tabskip=0pt\hrulefill}&\cr}
$$\vbox{\offinterlineskip\tabskip=0pt\halign{\hskip 1.0in
$#$\quad&\vrule#&\quad\hfil $\strut #$\hfil\quad &\vrule #&\quad\hfil
$#$\hfil\quad
&\vrule # &\quad\hfil $#$\hfil\quad &\vrule #&\quad\hfil $#$\hfil\quad
&\vrule
#&\quad\hfil
$#$\hfil\quad &\vrule{} \vrule #&\quad\hfil $#$\hfil\quad &\vrule #\cr
&\omit&\strut16&\omit&{10}&\omit&10'&\omit&S&\omit&S'
&\omit&\CW&\omit\cr
\tablerule
S_Q&&-1&&-2&&6&&0&&4&&-4&\cr
\tablerule
\noalign{\bigskip}
\noalign{\narrower\noindent{\bf Table 3:} Charges ($\in \BZ/20\BZ$) of
the
spacetime matter multiplets, and of the spacetime
superpotential, $\CW$, under $S_Q$, the ``quantum"
R-symmetry present at the \LG\ point.
}
 }}$$

80 $\bf{16}$s of $SO(10)$, 72 $10$s and 318 gauge singlets $S$
arise in the
$k=0$ sector of the \LG\ theory (for generic choices of the defining
data).  There are also 21 singlets $S'$ which arise in the $k=2$
twisted sector, and 2 $10$s of $SO(10)$ which
arise in the $k=4$ sector and are denoted as $10'$
in Table 3.  The detailed forms of the corresponding states can be found
in \DK\ \S4.1, and will not be important in what follows.

One sees immediately that the quantum symmetry \quantwo\ guarantees that
no terms of the form $f(S)$ or $f(S)S^\prime$ (where
$f$ is an arbitrary function of the 318 untwisted singlets)
can appear in the
spacetime superpotential $\CW$.  Therefore, as in \S2, one is guaranteed
that the corresponding 318 vertex operators are mutually integrable
moduli of the (0,2) theory.

\optional{
Again, it is important to check that the discrete R-symmetry is
nonanomalous. For the $Y_{W5;4,4}$ example discussed above, the anomaly is
$$(-2)\cdot c({\bf 45})+80\cdot (1)\cdot c({\bf 16})+\big(72\cdot (0)+2\cdot
(8)\big)\cdot c({\bf 10})=320=0\ {\rm mod}\  20$$
where $c({\bf 45})=16$, $c({\bf 16})=4$ and $c({\bf 10})=2$ for SO(10).
}

Of course, we had to $\it assume$ that one point in the moduli space of
this (0,2) \LG\ theory existed as a conformal theory to run this argument.
Making this assumption, we have proved that an entire 318 dimensional
moduli space of (0,2)
\LG\ theories exists.
The 318 singlets $S$ correspond to the $\bar Q_{+}$
nontrivial deformations of the Landau-Ginzburg superpotential.

We should go on at this point to analyse the spacetime superpotential for
the twisted-sector singlets $S'$, to see if there is a flat direction
which we can interpret as moving away from the \LG\ point in \Ka\ moduli
space. The analysis is, unfortunately, somewhat more complicated than the
case of the quintic. The flavour symmetry group is
$SU(7)\times SU(4)\times SU(2)$, and the polynomial coefficients lie in
three different irreducible representations of this group.
We hope to present this analysis elsewhere.

\newsec{The Renormalization Group}

The conclusions that we have been able to draw so far may seem  a little
an\ae mic. Using the quantum symmetry, we have been able to establish the
self-consistency of the (0,2) \LG\ theory, so we can be fairly confident that
the \LG\ theory, and the theory at infinite radius, are (0,2) superconformal in
the infrared. Moreover, these theories are clearly distinct. The former has a
discrete spectrum, whereas the latter has a continuous spectrum of states when
quantized on the circle.

 We have not, however, gotten very far in showing that
these theories remain superconformal as we deform in the
\Ka\ modulus. Nevertheless, with some plausible dynamical assumptions about the
behaviour of the linear
\sm s
\refs{\phases\DK}, we have enough information, for the simple case of models
with a single \Ka\ parameter, to prove that the whole phase diagram of the
linear
\sm\ is superconformal.

In the linear \sm, the worldsheet
superpotential is unrenormalized, even nonperturbatively\foot{The simplest
way to see this is to note that the flavour symmetries exploited in \S3
forbid any nontrivial renormalization of the $F$s.}. The issue which we need to
grapple with is the renormalization of the coefficient of the Fayet-Iliopoulos
D-term in the linear \sm\ action. This coefficient, in the infrared limit, is
nothing other than the
\Ka\  parameter.

Demanding that the one loop divergence
vanish imposes a condition on the sum of the scalar charges in the model,
which is easily satisfied in the models of interest
\refs{\phases\DK}. Beyond one loop, there is a nonrenormalization theorem
which says that the Fayet-Iliopoulos D-term is unrenormalized to all orders in
perturbation theory. What we need to worry about is whether the D-term is
renormalized {\it nonperturbatively}, say by gauge instantons.  If this
happens, we
might find ourselves in a situation where the running coupling $r(\mu)$ runs
off to infinity as we flow to the infrared. In that case, even though the the
linear \sm\ {\it seemed} to contain a continuously-variable \Ka\ parameter,
the infrared limit consists of only one point, the infinite radius theory.

More generally, the infrared limit might consist of several points, or it
might consist of the entire 2 real dimensional \Ka\ moduli space (what we
hope to prove).  Spacetime supersymmetry, which requires the moduli to form
chiral multiplets, forbids the remaining possibility -- a component of the \Ka\
moduli space of real dimension 1.\foot{The existence of a
1-dimensional moduli space of {\it non-supersymmetric} fixed points is
precluded by assumption 2) below.}

We now make some plausible assumptions about the behaviour of these theories.
\item{1)}If, along some RG trajectory, (0,2) supersymmetry is unbroken in the
infrared, then the central charge in the infrared limit is accurately given by
\ecbar. That is, we assume that the $U(1)$ R-symmetry that is present in these
models becomes the $U(1)$ current $\bar J$
in the right-moving N=2 algebra in the infrared
limit.

\item{2)}The only trajectories for which (0,2) supersymmetry is broken in the
infrared are those which flow to $r=\theta=0$, the ``phase transition point"
\phases. This is almost certainly true, because the Witten index is
well-defined and {\it nonzero} for all these theories {\it except} at
$r=\theta=0$
where the vacuum manifold becomes noncompact \phases.

We now examine the Zamolodchikov c-function \Zam\ as a function of the \Ka\
parameter for these theories.\foot{Equivalently, one could think of the
renormalization group flow as defining a vector field on the \Ka\ parameter
space, and apply the Lefschetz fixed-point theorem.}
Assume (counterfactually) that the critical
points of the c-function are isolated. The  (0,2) supersymmetric critical
points must, in fact, be local minima, since spacetime supersymmetry implies
the nonexistence of tachyons ( (0,2)-preserving relevant perturbations).
Assumption 1) says that these minima are all degenerate. We have established
the existence of at least two such minima: the infinite radius theory, and the
\LG\ theory. The only critical points which are {\it not} minima must
have (0,2) supersymmetry spontaneously broken. But, by assumption 2), the only
candidate for such a critical point is $r=\theta=0$. So we have a function
with two (or more) isolated minima, and one other critical point. This is
impossible. The  c-function is a Morse function on the \Ka\ parameter space
(topologically a sphere). The alternating sum of its critical points,
($\#$minima)$-(\#$ saddle-points)$+(\#$ maxima), must give the Euler
characteristic, and the number of critical points of index $i$ must be greater
than or equal to the $i^{th}$ Betti number. Under the above assumptions, these
Morse inequalities must be violated. Hence the hypothesis that
the critical points are isolated must be false, and the whole \Ka\ moduli space
is superconformal.

Note that this argument relied on the {\it crucial} fact that we had at least
two
minima. Had we not established that the \LG\ theory was superconformal, then
we could readily satisfy the Morse inequalities by
 giving the c-function one minimum (the infinite radius theory) and one
maximum ($r=\theta=0$).

This argument can be generalized to higher-dimensional \Ka\ moduli spaces,
provided we have sufficient control over the locus in the parameter space
on which supersymmetry may be broken. In the higher dimensional case,
the critical point sets of the c-function are no longer points, but manifolds
$M_j$,
so we need to use a simple generalization of the Morse inequalities due to
Bott \BottHomotopy,
 $$\eqalign{
  b_i(M)&\leq\sum_j b_{i-ind(M_j)}(M_j)\cr
 \chi(M)&=\sum_j(-1)^{ind(M_j)}\chi(M_j)\quad.\cr
 }$$

\newsec{Discussion}
Let us review what we have seen
in the previous sections:
\item{1)} There is strong evidence that the (0,2) Landau-Ginzburg
orbifolds of \DK\ have nontrivial infrared fixed points.  For certain
(0,2) deformations of (2,2) theories, the discussion of \S2 constitutes
a proof of this, and demonstrates that many (2,2) \LG\ theories are adjoined
to
much larger spaces of (0,2) \LG\ theories.
\item{2)} The (2,2) and (0,2) \LG\ theories often contain many
massless $E_{6}$ singlets which $a ~priori$ might become massive as one
deforms away from the \LG\ point.  Symmetry arguments like those of \S3
indicate that often, most of these $E_{6}$ singlets remain massless as
one leaves the \LG\ point.  These symmetry arguments also constrain the
singlet
$n$-point functions for low $n$.
\item{3)} For the case of one-parameter \Ka\ moduli spaces,
one can prove using simple topological arguments that the (0,2) moduli
spaces of 1) continue to exist at finite radius.
For (2,2) theories like the quintic, this indicates that the full (2,2)
moduli space is a submanifold of a much larger (0,2) moduli space.
Similar topological arguments may allow one to prove analogous
statements for theories with multi-dimensional \Ka\ moduli spaces, given
sufficient knowledge of the ``phase diagram.''

In the (0,2) context, having an extended chiral
algebra with (2,2) superconformal supersymmetry is a very rare situation and
requires highly nongeneric defining data. Still, it would not be surprising
if the locus in (0,2) moduli space on which this occurred had several disjoint
components. In this situation,  a picture like \tfig\figtwo\ could arise.
A single (0,2) moduli space could connect different (2,2) moduli spaces,
whose large-radius limits correspond to {\it different} Calabi-Yau manifolds,
which are perhaps not even birationally equivalent.

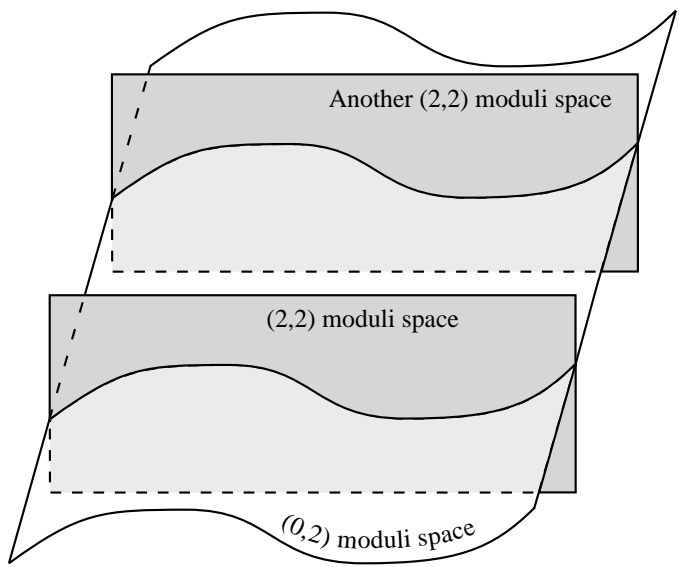
\ifigure\figtwo{Two different (2,2) moduli spaces characterized by
the same number of generations and antigenerations
arising as different enhanced
symmetry points in a larger (0,2) moduli space.}{modtwo.eps}{3.0}

As we have seen, the number of massless generations and antigenerations
do not change as we move about in the particular
(0,2) moduli space we have discovered here. Thus the two
different Calabi-Yau manifolds in the above picture must have the same Hodge
numbers.

Equally possible is that there are other ``special points" in (2,2) moduli
space which are multicritical, in that extra (0,2) flat directions appear
there. In that case, the picture of the moduli space becomes even more
complicated than \figtwo\ (exceeding our artistic abilities to depict
it).  One
can pass from a (2,2) theory to a (0,2) theory to another (2,2) theory to
yet another (0,2) theory to $\dots$ In this way, one might conjecture that all
Calabi-Yau manifolds with the same Hodge numbers are continuously connected to
each other.

Of course, we have only considered turning on VEVs for gauge-singlets. It is
also possible (and, indeed, cases are known \miracles) that there are F- and
D-flat perturbations which {\it break} the $E_{6}$
gauge symmetry. In this case, the
only invariant we expect to be preserved as one moves about in the
(0,2) moduli space is the
difference between the number of generations and antigenerations.\foot{Which
is given by ${1\over 2}\chi$ for (2,2) Calabi-Yau theories and ${1\over
2}c_{3}(E)$ for (0,2) Calabi-Yau theories with vacuum gauge bundle $E$.}

\bigbreak\bigskip\bigskip\centerline{{\bf Acknowledgments}}\nobreak
\frenchspacing{
This work was supported by NSF grant PHY90-21984
and by the A.P. Sloan Foundation. JD would like to thank the Theory
Group at Rutgers University for its hospitality during the early stages of
this work. We would like to thank E. Witten, S. Shenker, and N. Seiberg for
illuminating conversations.
}

\listrefs
\end